\documentclass[aps,prd,superscriptaddress,twocolumn,secnumarabic,nofootinbib]{revtex4-1}
\pdfoutput=1
\usepackage[utf8]{inputenc}
\usepackage[normalem]{ulem} 
\usepackage[english]{babel}
\usepackage{tabularx}
\usepackage{array}
\usepackage{graphics}
\usepackage[pdftex]{graphicx}
\usepackage{mathtools}
\usepackage{amssymb}
\usepackage{amsmath}
\usepackage{slashed}
\usepackage{braket}

\usepackage[usenames,dvipsnames]{xcolor}
\usepackage{bm}

\usepackage{url}
\usepackage{hyperref}
\hypersetup{colorlinks,citecolor=nicegreen,linkcolor=niceblue}
\hypersetup{colorlinks=true}

\newcommand{\tev}{{\rm TeV}}
\newcommand{\gev}{{\rm GeV}}
\newcommand{\mev}{{\rm MeV}}

\newcommand{\nn}{\nonumber}

\newcommand {\E}[1]{\times 10^{#1}}	
\newcommand {\e}[1]{\mathrm{\,#1}}       
\newcommand{\mc}[1]{\mathcal{#1}}
\newcommand{\mrm}[1]{\mathrm{#1}}

\newcommand{\re}[0]{\mrm{Re}}
\newcommand{\im}[0]{\mrm{Im}}

\newcommand{\acp}[0]{\mathcal{A}_\mathrm{CP}}

\newcommand{\naiveT}{\mathrm{T}_\mathrm{N}}

\definecolor{niceblue}{rgb}{0.15,0.15,0.6}
\definecolor{nicegreen}{rgb}{0.1,0.5,0.1}
\definecolor{Red}{rgb}{1.,0.,0.}

\definecolor{Green}{rgb}{0.2,.7,0.2}

\newcommand{\rkkst}[0]{R_{K^{(*)}}}
\newcommand{\rk}[0]{R_K}
\newcommand{\rkst}[0]{R_{K^*}}

\newcommand{\qlq}[1]{Q_{lq}^{(#1)}}
\newcommand{\qqq}[1]{Q_{qq}^{(#1)}}

\newcommand{\cqq}[1]{C_{qq}^{(#1)}}

\begin{document}
\title{LFU and CP violation with $S_3$}

\author{Nejc Ko\v snik} \email[Electronic address:]{\ nejc.kosnik@ijs.si}
\affiliation{Department of Physics, University of Ljubljana, Jadranska 19, 1000 Ljubljana, Slovenia}
\affiliation{Jo\v zef Stefan Institute, Jamova 39, P.\ O.\ Box 3000, 1001
  Ljubljana, Slovenia}

\author{Aleks Smolkovi\v c} \email[Electronic address:]{\ aleks.smolkovic@ijs.si}
\affiliation{Jo\v zef Stefan Institute, Jamova 39, P.\ O.\ Box 3000, 1001
  Ljubljana, Slovenia}

\begin{abstract}
  We introduce the CP violating scalar leptoquark $S_3$ to explain the measured values of the lepton universality ratios $R_{K^{(*)}}$. We derive constraints on the CP-even and CP-odd components of the leptoquark Yukawa couplings stemming from effects in $b \to s \mu \mu$ and $B_s$ mixing. For the $b \to s \mu \mu$ processes we impose $R_{K^{(*)}}$, $\mathcal{B}(B_s \to \mu^+ \mu^-)$, as well as CP-sensitive angular asymmetries $A_{7,8,9}$, whereas in the $B_s$ mixing sector $\Delta M_s$ and $S_{\psi\phi}$ are considered. Combining the constraints within the $S_3$ model reveals that a large CP phase with a definite sign is perfectly viable for a leptoquark of mass below a few TeV. For larger mass of the $S_3$ leptoquark the CP phase is suppressed due to the observables pertaining to the $B_s$ system. We provide predictions of direct and mixing-induced CP asymmetries in $B \to K \mu \mu$ that could reveal the presence of the novel CP phase.
\end{abstract}

\maketitle

\section{Introduction}
The flavour structure of the Yukawa sector is probably the least
understood aspect of the Standard
Model~(SM).
In regard to the quark Yukawa couplings, numerous experiments have confirmed the general validity of the Cabibbo-Kobayashi-Maskawa~(CKM) paradigm, which predicts all quark flavour transitions in terms of four
parameters. The CKM matrix has small flavour violating elements and a
unique phase that drives all CP-violating
quantities. Furthermore, flavour changing neutral quark currents (FCNC) are suppressed due to only occurring at higher order in perturbation theory, as well as due to the unitarity of the CKM matrix via the Glashow-Iliopoulos-Maiani mechanism. On the contrary, in
the lepton sector, the Pontecorvo-Maki-Nakagawa-Sakata mixing
matrix exhibits large flavour mixing, however the smallness of neutrino
masses renders this effect unobservable in experiments blind to
neutrino flavours. 
Consequently, the flavour of charged leptons is conserved and couplings of leptons to gauge bosons are lepton flavour universal~(LFU), whereas
differing masses of leptons explicitly break LFU. Thus, in a LFU ratio of two processes, which are related by a lepton flavour rotation, systematic errors largely cancel, provided there is a large overlap in the phase space~\cite{hep-ph/0310219,1408.1627}.

The LFU predictions have been tested in various processes on energy scales ranging
from kaon decays, weak boson decays $Z \to \ell \ell$ at LEP~\cite{1809.06229}, to
highest energy tests at the Large Hadron Collider~(LHC)~\cite{CMS:2021ctt}. In the last decade, the LHCb experiment presented measurements of LFU-sensitive ratios $\rkkst \equiv {\Gamma' (B\to K^{(\ast)} \mu\mu)}/{\Gamma'(B\to K^{(\ast)} e e)}$, where $\Gamma'$ stands for the partial width in the region $q^2 \in [1.1,6]\,\mrm{GeV}^2$, and found
  \begin{equation}
  \label{eq:RKKstExp}
  \begin{split}
        \rk &= 0.846^{+0.042+0.013}_{-0.039-0.012}\quad\text{\cite{Aaij:2014ora,Aaij:2019wad,Lancierini:2021aek}},\\
    \rkst &=0.69^{+0.11+0.05}_{-0.07-0.05}\quad\text{\cite{Aaij:2017vbb}}.
  \end{split}    
  \end{equation}
The ratios are $3.1$ and $2.6\,\sigma$ below their SM predictions,
$R_{K^{(\ast)}}=1.00(1)$~\cite{Bordone:2016gaq}. Driven by FCNC, this process is GIM suppressed, which allows potential
New Physics~(NP) contributions to stand out. In the effective
Hamiltonian description the presence of a NP effective operator with
left-handed fermions
$\mc{O}_9 - \mc{O}_{10} \propto (\bar s_L \gamma^\mu b_L)(\bar \mu_L
\gamma_\mu \mu_L)$ is in good agreement
with measurements in Eq.~\eqref{eq:RKKstExp}, as well as with the global set of
$b\to s \mu \mu$ observables~\cite{2104.10058,Altmannshofer:2021qrr}.
Scalar or vector leptoquarks (LQs) at the TeV scale can naturally generate
such effective operators at tree-level, see e.g.~\cite{Angelescu:2021lln}. The discrepancies in Eq.~\eqref{eq:RKKstExp} pull the value of the NP Wilson coefficient $\delta C_9 = -\delta C_{10}$ to negative real values and possibly large CP-violating imaginary parts, which are allowed by global analyses~\cite{Altmannshofer:2021qrr, Carvunis:2021jga}. Such NP CP-violating phases lead to enhanced direct CP-asymmetries in $B \to K^{(*)} \mu^+ \mu^-$ decays~\cite{Becirevic:2020ssj}.

Large $\im(\delta C_9)$ imprints its CP violating effects on processes connected to $b \to s \mu \mu$ via electroweak mixing. Such a connection between the $\rkkst$ anomalies and the $B_s$ mixing observables has been studied in Refs.~\cite{1703.09247,1712.06572,1909.11087}. In $Z'$ and LQ models $\delta C_9$ and $B_s-\bar B_s$ are concurrently generated, however the correlations between the two processes are quite different in the two models. The CP violating aspects of $B_s$ mixing and their relation to $\delta C_9$ have been studied in~\cite{1703.09247} in the context of $Z'$ models. Connection between the $\rkkst$ anomalies and $B_s$ mixing frequency $\Delta m_s$ has already been studied in Refs.~\cite{1712.06572,1909.11087}.

In this Letter we consider the phenomenological consequences of $\im(\delta C_9)$, generated by a concrete model with a scalar LQ $S_3=(\mathbf{\overline{3}},\mathbf{3},1)$, and paying particular attention to the effects in $B_s$ mixing. This LQ naturally realises the left-handed scenario $\delta C_9 = -\delta C_{10}$ since the gauge quantum numbers allow only Yukawa couplings to fermion doublets~\cite{1408.1627}. This LQ state was also considered in attempts to address $\rkkst$ and $R_{D^{(*)}}$ simultaneously~\cite{1703.09226,Becirevic:2018afm,Crivellin:2019dwb,2008.09548,2103.12504} as well as in broader context including dark sectors~\cite{Choi:2018stw,DEramo:2020sqv}.

In Sec.~\ref{sec:matching} we perform matching onto appropriate operators for $\delta C_9$ and the $B_s$ mixing coefficient $C^{LL}_{bs}$ where we allow for arbitrary complex Yukawa couplings of the LQ. In Sec.~\ref{sec:constr} we derive constraints stemming from CP (non-)conserving observables in $b\to s \mu \mu$ and $B_s-\bar B_s$ mixing.
Next, we show in Sec.~\ref{sec:combination} to what extent the CP conserving and violating $b\to s \mu \mu$ and $B_s$ mixing observables $\Delta m_s$ and $S_{\psi\phi}$ restrict the allowed values of $\delta C_9$. We predict possible future signatures of the leptoquark CP phase, and finally conclude in Sec.~\ref{sec:conc}.

\section{Matching}
\label{sec:matching}
\subsection{Dimension-6 operators from the $S_3$ leptoquark}
Here we present the effects of the $S_3$ LQ via the matching onto the left-handed four-fermion operators of the SMEFT~\cite{Grzadkowski:2010es,1308.2627,1310.4838,1312.2014}. 
We pick only those LQ Yukawa couplings at the matching scale that are necessary to induce the operators with the flavor structure $(\bar s b)(\bar \mu \mu)$.

The starting point is the Lagrangian of the $S_3$ leptoquark~\cite{1603.04993}
\begin{equation}
  \label{eq:LS3}
  \mc{L} = |D_\mu S_3|^2 - m_{S_3}^2 |S_3|^2 + y_{ij} \overline{Q_i^C} (i\tau^2 \tau^I)  L_j \, S_3^I,
\end{equation}
where the index $I=1,2,3$ runs over the weak isospin generators. We assume that the quark flavour index $i=d,s,b$ refers to the down-quark mass eigenstates, forcing the CKM matrix $V$ to appear alongside up-type quark mass eigenstates, $Q_i = (V^*_{ki} u_{L,k}, d_{L,i})^T$. We assume that diquark couplings, allowed by the SM charges of $S_3$, are forbidden by a suitable symmetry in order not to cause rapid proton decay. We have neglected neutrino masses and used the index $j$ to refer to charged lepton mass eigenstates. Our basic assumption is that at scale $m_{S_3}$ only the Yukawa couplings $y_{b \mu}$ and $y_{s \mu}$ are non-zero to accommodate $\rkkst$ measurements, and that these couplings can take complex values.
We match onto the SMEFT Lagrangian, defined as $\mc{L}_\mrm{dim-6} =\sum_i C_i Q_i$, and find that the following semileptonic operators are generated at tree-level:
\begin{align}
  \label{eq:Qlq}
  \qlq{1} &= (\bar L_p \gamma^\mu L_r)(\bar Q_s \gamma_\mu Q_t),\\
            \qlq{3} &= (\bar L_p \gamma^\mu \tau^I L_r)(\bar Q_s \gamma_\mu \tau^I Q_t).
\end{align}
The flavour indices $(pqrst) = (\mu\mu s b)$ are fixed by the LQ Yukawa couplings $y_{b\mu}$ and $y_{s\mu}$ that enter the corresponding Wilson coefficients
\begin{equation}
  \label{eq:Clq}
  C_{lq}^{(1)} = \frac{3 y_{b\mu}y_{s\mu}^\ast}{4m_{S_3}^2},\qquad C_{lq}^{(3)} = \frac{y_{b\mu}y_{s\mu}^\ast}{4m_{S_3}^2},
\end{equation}
On the other hand, at one-loop level we get four-quark operators $\qqq{1,3}$
\begin{align}
  \label{eq:Qqq}
  \qqq{1} &=  (\bar Q_p \gamma^\mu Q_r)(\bar Q_s \gamma_\mu Q_t),\\
  \qqq{3} &= (\bar Q_p \gamma^\mu \tau^I Q_r)(\bar Q_s \gamma_\mu \tau^I Q_t).
\end{align}
with indices $(pqrst) = (sbsb)$, thus contributing at low scales to $B_s-\bar B_s$ mixing. The respective Wilson coefficients read
\begin{equation}
  \label{eq:Cqq}
    C_{qq}^{(1)} = -\frac{9 (y_{b\mu}y_{s\mu}^\ast)^2}{256 \pi^2 m_{S_3}^2},\qquad C_{qq}^{(3)} = -\frac{(y_{b\mu}y_{s\mu}^\ast)^2}{256 \pi^2 m_{S_3}^2}.
\end{equation}
Our matching results agree with Ref.~\cite{2003.12525}.

\subsection{From SMEFT to WET}
The semileptonic operators $\qlq{1,3}$ at the matching scale naturally match onto the weak effective theory in the broken phase of the electroweak symmetry. Low-energy effects in $b \to s \mu^+ \mu^-$ are parameterized by the following operators
\begin{equation}
	\label{eq:EffLagbsmumu}
	\mathcal{H}^{b\to s \ell \ell}_{\mathrm{eff}} = -\dfrac{4 G_F  V_{tb}V_{ts}^\ast}{\sqrt{2}}  \sum_{i=7,9,10} C_i(\mu)\mathcal{O}_i(\mu)
\end{equation}
and we set the scale to $\mu = m_b$. Relevant modification of the effective Wilson coefficients valuesin the SM~\cite{hep-ph/9311345,0811.1214} are $C_{9,10} = C_{9,10}^\mathrm{SM} \pm \delta C_{9}$. Note that due to the LQ flavour structure, only operators with muons are modified. The relevant operators are
\begin{align}
	\label{eq:O7910}
	\mc{O}_7  &=  \dfrac{e m_b}{4\pi} \, (\bar{s}_R \sigma_{\mu\nu}  b_R ) F^{\mu\nu}\,,\\
	\mathcal{O}_{9(10)} &= \dfrac{e^2}{(4\pi)^2} \,(\bar{s}_L\gamma_\nu  b_L ) (\bar\mu\gamma^\nu (\gamma^5)\mu )\,.
\end{align}
The QCD and QED renormalization group running below the weak scale of such quark current operators is negligible~\cite{1706.00410}. Thus we can read off the semileptonic coefficients at scale $\mu = m_b$ from the expressions in Eq.~\eqref{eq:Clq}:
\begin{equation}
	\label{eq:C9C10}
\delta C_9 = -\delta C_{10} = \frac{\pi y_{b\mu} y_{s \mu}^*}{\sqrt{2} G_F V_{tb} V_{ts}^* \alpha_\mrm{em} m_{S_3}^2}.
\end{equation}
At loop level also the coefficient of the dipole operator gets modified by $\delta C_7$, however the contribution is strongly suppressed by a loop factor and $\alpha$, i.e. $\delta C_{7} = \alpha/(8\pi)\,\delta C_9$.

The coefficients of the four-quark operators~\eqref{eq:Cqq}, on the other hand, match onto the $\Delta B = \Delta S = 2$ effective Lagrangian with exclusively left-handed quarks~\cite{1909.11087}
\begin{equation}
	\label{eq:LBsmix}
	\mc{L}_{bs} = -\frac{4 G_F}{\sqrt{2}} (V_{tb} V_{ts}^\ast)^2 C^{LL}_{bs}(\mu)\,(\bar s_L \gamma^\mu b_L)^2,
\end{equation}
with the modification of the Wilson coefficient $C^{LL}_{bs} = C^{LL (\mathrm{SM})}_{bs} + \delta C^{LL}_{bs}$. The SM part reads~\cite{1511.09466,1909.11087}
\begin{align}
\label{eq:CLLSM}
C^{LL(\mathrm{SM})}_{bs}(m_b) &= \hat\eta_B \frac{m_W^2 S_0(\bar m_t^2/m_W^2)}{16\pi^2 v^2} \\
&= (1.310 \pm 0.010)\times 10^{-3},\nn
\end{align}
and contains the $\mu$-dependent QCD renormalization group factor
$\hat \eta_B = 0.84$~\cite{1511.09466,Buras:1990fn} due to running to
the $\mu = m_b$ scale. For the LQ contribution we match and run the
coefficients $\cqq{1,3}$ in Eq.~\eqref{eq:Cqq} to
\begin{equation}
	\label{eq:CbsLL}
	\delta C^{LL}_{bs}(\mu = m_b) = \eta^{6/23}\frac{5 (y_{b\mu} y_{s \mu}^*)^2}{256 \sqrt{2} \pi^2 m_{S_3}^2  G_F (V_{tb} V_{ts}^*)^2},
\end{equation}
where $\eta = \alpha_s(m_{S_3})/\alpha_s(m_b)$ at leading order,
neglecting the top quark threshold
effect~\cite{Buras:1990fn,1712.06572}.

\section{CP-even and odd constraints}
\label{sec:constr}
In this Section we present the relevant observables and derive constraints on low energy effective interactions~\eqref{eq:EffLagbsmumu} and \eqref{eq:LBsmix}. We have checked that additional CP-odd signatures of the model, e.g. electric dipole moments of $c$-quark or muon, also generated by the SMEFT interaction~\eqref{eq:Clq}, are well beyond current upper experimental limits and we do not discuss them here.

\subsection{$b\to s \mu\mu$ constraints}
In order to predict $R_{K^{(*)}}$, we employ the numerical formulae~\cite{Bordone:2016gaq}, valid for the $\delta C_9 = -\delta C_{10}$ scenario in the region $q^2 \in [1.1,6]\,\mrm{GeV}^2$:
\begin{equation}
  \label{eq:RKC9C10}
  \begin{split}
      R_K &= 1.00 + 0.48\, \re(\delta C_9)\\
      &\quad- 8.3\E{-3}\, \im(\delta C_9) 
        + 0.057 \,|\delta C_9|^2,\\
      R_{K^*} &= 1.00 + 0.47 \,\re(\delta C_9)\\
       &\quad- 8.3\E{-3}\, \im(\delta C_9)
      + 0.064\,|\delta C_9|^2.
    \end{split}
\end{equation}
Although $R_{K^{(*)}}$ is defined as a ratio of CP-averaged decays widths the linear term $\im(\delta C_9)$ arises due to interference with the small CPV phase in $V_{tb} V_{ts}^*$.
The branching ratio of the $B_s$ muonic decay has been re-measured recently by LHCb~\cite{LHCb:2021awg} and was combined with ATLAS~\cite{1812.03017} and CMS~\cite{1910.12127} in the world average that reads $\mc{B}(B_s \to \mu\mu) = (2.85^{+0.34}_{-0.31})\E{-9}$~\cite{2104.08921}. The theoretical expression reads
\begin{align}
\label{eq:BS}
\mathcal{B}(B_s\to \mu^+\mu^-) &= \frac{\tau_{B_s}}{1-y_s}\dfrac{\alpha^2 G_F^2 m_{B_s}}{16 \pi^3}  \left| V_{tb}V_{ts}^\ast \right|^2  m_\mu^2   \nonumber \\
& \quad\times \sqrt{1-\frac{4 m_\mu^2}{m_{B_s}^2}} \left|C_{10}\right|^2 f_{B_s}^2,
\end{align}
where $f_{B_s}= (230.3\pm 1.3)\,\mev$~\cite{Aoki:2019cca}, and one must also account for the effect of $B_s$ oscillations in the time-integrated measurement of the branching fraction~\cite{1204.1735}. In Ref.~\cite{1908.07011} the authors computed power-enhanced QED corrections to the rate and found for the SM prediction $\mathcal{B}(B_s\to \mu^+\mu^-)^\mrm{SM} = (3.66\pm 0.14)\E{-9}$. In our analysis we rescale this SM prediction as $\mathcal{B}(B_s\to \mu^+\mu^-)^\mrm{th} = \left|1+ \delta C_{10}/C_{10}^\mrm{SM}\right|^2\mathcal{B}(B_s\to \mu^+\mu^-)^\mrm{SM}$, where we take for the SM value of the axial current coefficient $C_{10}^\mrm{SM} = -4.103$.

As we allow for complex Wilson coefficients $\delta C_9 \in \mathbb{C}$, we also consider the $\naiveT$-odd\footnote{The naive time reversal $\naiveT$ reverses the momenta and spins of particles, not to be confused with $\mathrm{T}$, which additionally exchanges the initial and final states.} CP-odd observables $A_{7,8,9}$, which are sensitive to CP-violating weak phases even in absence of strong phases~\cite{0805.2525}. We use the latest LHCb measurements of $A_{7,8,9}$ in $q^2 \in [1.1,6]\,\mrm{GeV}^2$, provided in Ref.~\cite{1512.04442}. As for the theoretical predictions, we rely on the $\texttt{flavio}$ package~\cite{1810.08132} and check that the constraints agree with the ones provided in Refs.~\cite{Altmannshofer:2021qrr,Carvunis:2021jga}.

\subsection{$B_s-\overline{B}_s$ mixing constraints}
\begin{figure}[t]
  \centering 
  \includegraphics[scale=0.55]{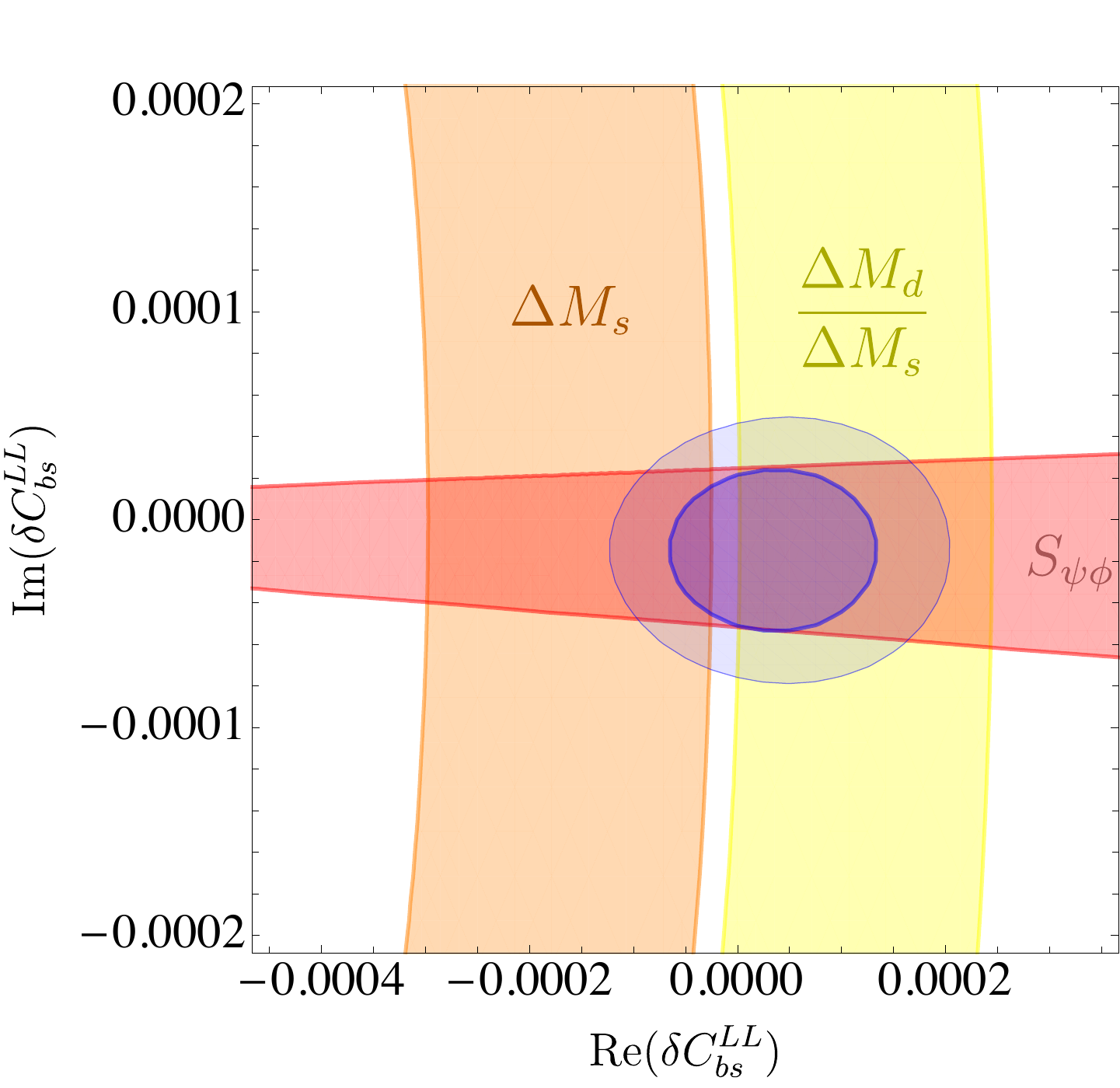}
  \caption{The $1\sigma$ constraints on $\delta C_{bs}^{LL}$ due to $B_s -\bar B_s$ mixing
    observables. The dark and light blue oval regions show the 1 and 2$\sigma$ fitted regions, respectively.}
	\label{fig:Bsmixing}
\end{figure}
Possible contributions of NP to $C_{bs}^{LL}$ in Eq.\eqref{eq:LBsmix} will be susceptible to constraints from $B_s-\overline{B}_s$ mixing. The impact on the mass difference $\Delta M_s$ can be parametrized as
\begin{equation}
\Delta M_s^{\mathrm{SM}+ \mathrm{NP}}  = \Delta M_s^{\mathrm{SM}}\left|1 + \frac{\delta C^{LL}_{bs}(m_b)}{C^{LL(\mathrm{SM})}_{bs}(m_b)} \right|,
\end{equation}
where
\begin{equation}
  \label{eq:dmsSM}
  \Delta M_s^{\mathrm{SM}} = \frac{4\sqrt{2}}{3} G_F m_{B_s} C_{bs}^{LL(\mrm{SM})} |V_{tb} V_{ts}^*|^2 \,(f_{B_s} \sqrt{B_{B_s}})^2,
\end{equation}
where $C_{bs}^{LL(\mrm{SM})}$ is given by Eq.~\eqref{eq:CLLSM}. For the combination of the nonperturbative parameters entering the hadronic mixing amplitude we take the world average of lattice computations, $f_{B_s} \sqrt{\hat B_{B_s}} = 0.274(8)\e{GeV}$, prepared by the FLAG group~\cite{Aoki:2019cca}. To translate to scale-dependent $B_{B_s}$ we use the relation $\hat B_{B_s}/B_{B_s}(\mu_b) = 1.519$~\cite{1909.11087}.
The CKM elements entering the SM prediction are taken from the
CKMfitter tree-level fit of Summer
'18~\cite{hep-ph/0406184}\footnote{The results do not change
  significantly if we use NP fit of CKM
  parameters~\cite{2006.04824}.},
$V_{td} V_{ts}^* = -0.0424^{+0.0015}_{-0.0004}$. Using these inputs we
find
\begin{equation}
  \label{eq:dmsSMnumber}
  \Delta M_s^\mrm{SM} = (20.9^{+1.4}_{-1.7})\e{ps}^{-1},
\end{equation}
which is somewhat above the world experimental average $\Delta M_s^\mrm{HFLAV} = (17.74\pm 0.02)\e{ps}^{-1}$~\cite{HFLAV:2019otj}. On the other hand, if we consider the ratio $\Delta M_d/\Delta M_s$, where $\Delta M_d$ is the mixing frequency of the $B_d -\bar B_d$ system, the theoretical systematic errors are different. In our model $\Delta M_d = \Delta M_d^\mrm{SM}$ and thus the prediction reads
\begin{align}
  \label{eq:dmdOverdms}
    \left(\frac{\Delta M_d}{\Delta M_s}\right) = \left(\frac{\Delta M_d}{\Delta M_s}\right)^\mrm{SM}\,\left|1 + \frac{\delta C^{LL}_{bs}(m_b)}{C^{LL(\mathrm{SM})}_{bs}(m_b)} \right|^{-1},
\end{align}
where the SM value now depends on the nonperturbative parameter $\xi = \frac{f_{B_s} \sqrt{B_{B_s}}}{f_{B_d} \sqrt{B_{B_d}}} = 1.206(17)$~\cite{Aoki:2019cca} and the combination $|V_{td}/V_{ts}| = 0.215(5)$ obtained from the CKMfitter tree-level fit~\cite{hep-ph/0406184}:
 \begin{equation}
   \label{eq:dmdOverdmsSM}
   \left(\frac{\Delta M_d}{\Delta M_s}\right)^\mrm{SM} = \frac{1}{\xi^2} \left|\frac{V_{td}}{V_{ts}} \right|^2 \,\frac{M_{B_d}}{M_{B_s}}= 0.0311^{+0.0018}_{-0.0017},
 \end{equation}
 which is slightly above the world average $(\Delta M_d/\Delta M_s)^\mrm{HFLAV} = 0.02855 \pm 0.00011$~\cite{HFLAV:2019otj} and suggests a positive contribution of NP to $\Delta M_s$, contrary to what we would conclude from the $\Delta M_s$ observable alone. To overcome this quite ambiguous situation we take both observables into account to derive constraints on complex $\delta C_{bs}^{LL}$. We plot the two constraints in the complex plane of $\delta C_{bs}^{LL}$ in Fig.~\ref{fig:Bsmixing} and observe that they are insensitive to $\im(\delta C_{bs}^{LL})$.

In addition, the CP asymmetry $S_{\psi\phi}$ from the interference between $B_s$ mixing and the decay $B_s\to J/\psi \phi$~\cite{1511.09466} can be used to constrain the CP-violating $\im(\delta C_{bs}^{LL})$.
 The impact of NP can be parametrized as
\begin{equation}
S_{\psi\phi} \equiv \sin(-2 \beta_s + \delta\phi),
\end{equation}
with
\begin{equation}
\delta\phi = \mathrm{Arg}\left(1 + \frac{\delta C^{LL}_{bs}(m_b)}{C^{LL(\mathrm{SM})}_{bs}(m_b)} \right).
\end{equation}
We use the latest HFLAV experimental result of $S_{\psi \phi} = -0.050 \pm 0.019$~\cite{HFLAV:2019otj} in our analysis and the SM value $\beta _s = 0.0198^{+0.0012}_{-0.0009}$, again determined from the tree-level fit of CKMfitter.
Fig.~\ref{fig:Bsmixing} shows constraints in the $\delta C_{bs}^{LL}$ complex plane from the aforementioned $B_s$ mixing observables, where the $S_{\psi \phi}$ cuts away significant part of the parameter space with large $\im(\delta C_{bs}^{LL})$. The combined fit to all three $B_s-\bar B_s$ observables is shown in blue. We will employ this region in the next Section where we will interpret these constraints in the  $S_3$ LQ model.

\section{Implications for the $S_3$ model}
\label{sec:combination}
In the context of the $S_3$ LQ model the $b\to s \mu\mu$ and $B_s-\bar B_s$ processes, considered in the preceding Section, are in one-to-one correspondence as they are determined by the same couplings. The relation between the two effective interactions that is central to our analysis is the following:
\begin{equation}
\label{eq:CbsLL-C9}
\delta C_{bs}^{LL} = \eta^{6/23}\frac{5 G_F \alpha_\mathrm{em}^2}{128 \sqrt{2} \pi^4}(\delta C_9)^2 m_{S_3}^2.
\end{equation}
Here $\delta C_9$ is given in
Eq.~\eqref{eq:C9C10}. For fixed $\delta C_9$ the effect in $B_s - \bar B_s$ is
increasingly pronounced for larger $m_{S_3}$.  Another surprising
feature is that $\delta C_9 \in \mathbb{R}$ will always increase
$\Delta M_s$, while $\Delta M_s < \Delta M_s^\mrm{SM}$ can only be
attained by nonzero $\im(\delta C_9)$ (See also Ref.~\cite{
1909.11087}). The CPV constraint
$S_{\psi\phi}$ is not sensitive directly to CPV component
$\im(\delta C_9)$ but rather to the cross term
$\re(\delta C_9)\,\im(\delta C_9)$ between CP-conserving and
CP-violating LQ couplings entering $\rkkst$.

\subsection{Combined constraints}
\begin{figure}[b]
	\centering 
	\includegraphics[scale=0.55]{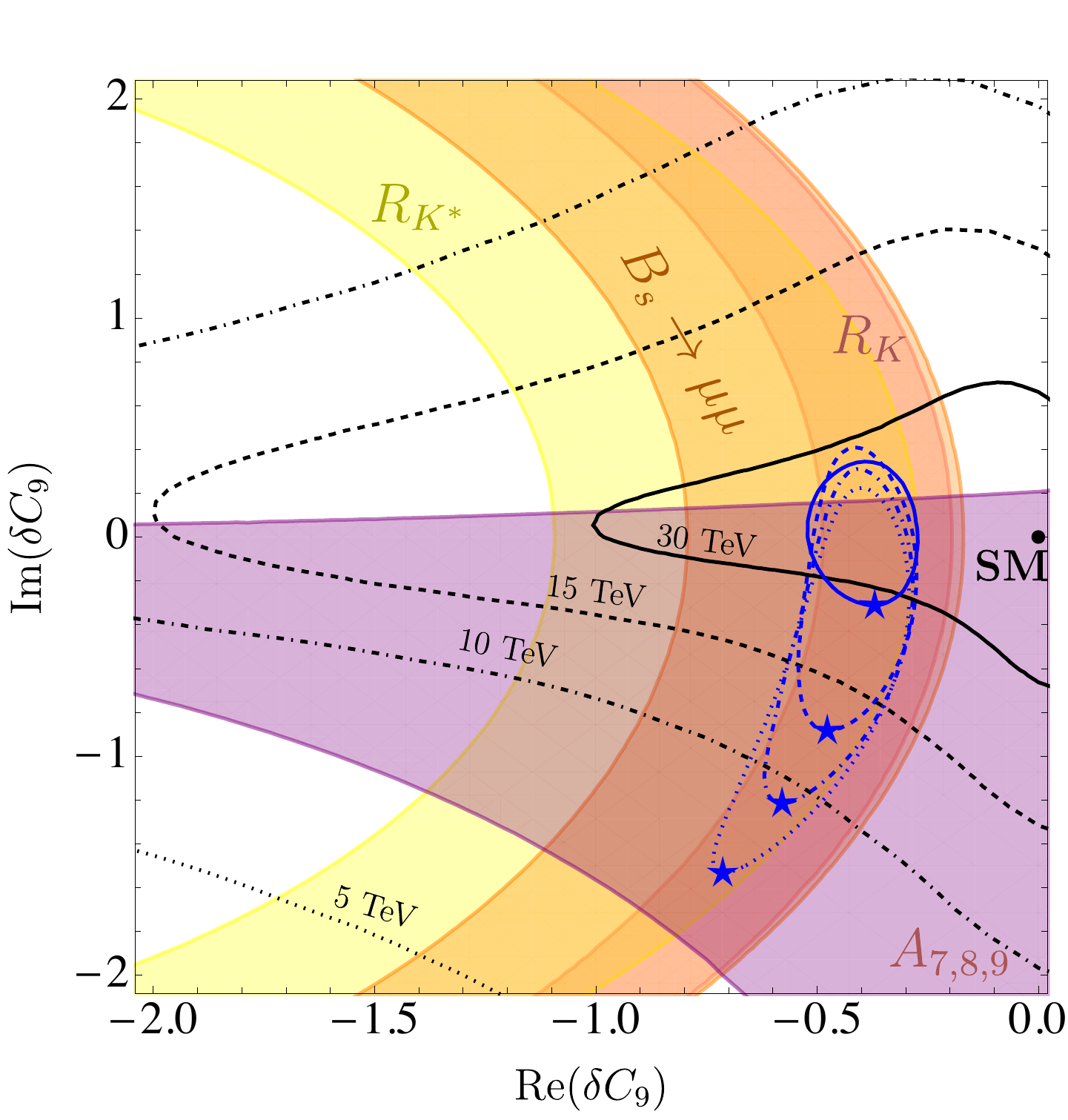}
	\caption{$1\sigma$ constraints on the $\delta C_9$ complex plane: model independent constraints from $R_{K^{(\ast)}}$, $\mathcal{B}(B_s>\mu\mu)$ and CP-odd angular coefficients $A_{7,8,9}$ from $B\to K^\ast \mu \mu$, as well as model dependent constraints from $B_s$ mixing for $4$ benchmark values of $m_{S_3}$, written on the black contours. The blue contours denote the best fit regions, with the stars showing the points of maximal allowed $\im(\delta C_9)$ for each benchmark.}
	\label{fig:C9S3}
\end{figure}
We now combine the $b\to s \mu \mu$ and $B_s$ mixing constraints, discussed in the previous Section, in the complex $\delta C_9$ plane shown in Fig.~\ref{fig:C9S3}. The bounds from $R_K$, $R_{K^\ast}$, $\mathcal{B}(B_s \to \mu \mu)$ and $A_{7,8,9}$ are all independent of $m_{S_3}$ (at the expense of varying LQ Yukawa couplings). It is worth emphasizing that the bound from $A_{7,8,9}$ has a preferred direction in $\im(\delta C_9)$. As expected, the bounds from $B_s$ mixing constraints are more stringent for higher $S_3$ masses. We show the constraints from the combined fit (Fig.~\ref{fig:Bsmixing}), translated via Eq.~\eqref{eq:CbsLL-C9}, in black contours for $m_{S_3} \in \{5,10,15,30\}~\mathrm{TeV}$. The combined fit of all the considered constraints is shown in blue contours, again for the same considered masses of $S_3$. With blue stars we denote benchmark points, which are defined so as to attain maximal value of $\im(\delta C_9)$ at $1\sigma$ (for given $m_{S_3}$).

\subsection{Predictions of CPV observables}
\begin{figure}[t]
	\centering 
	\includegraphics[scale=0.55]{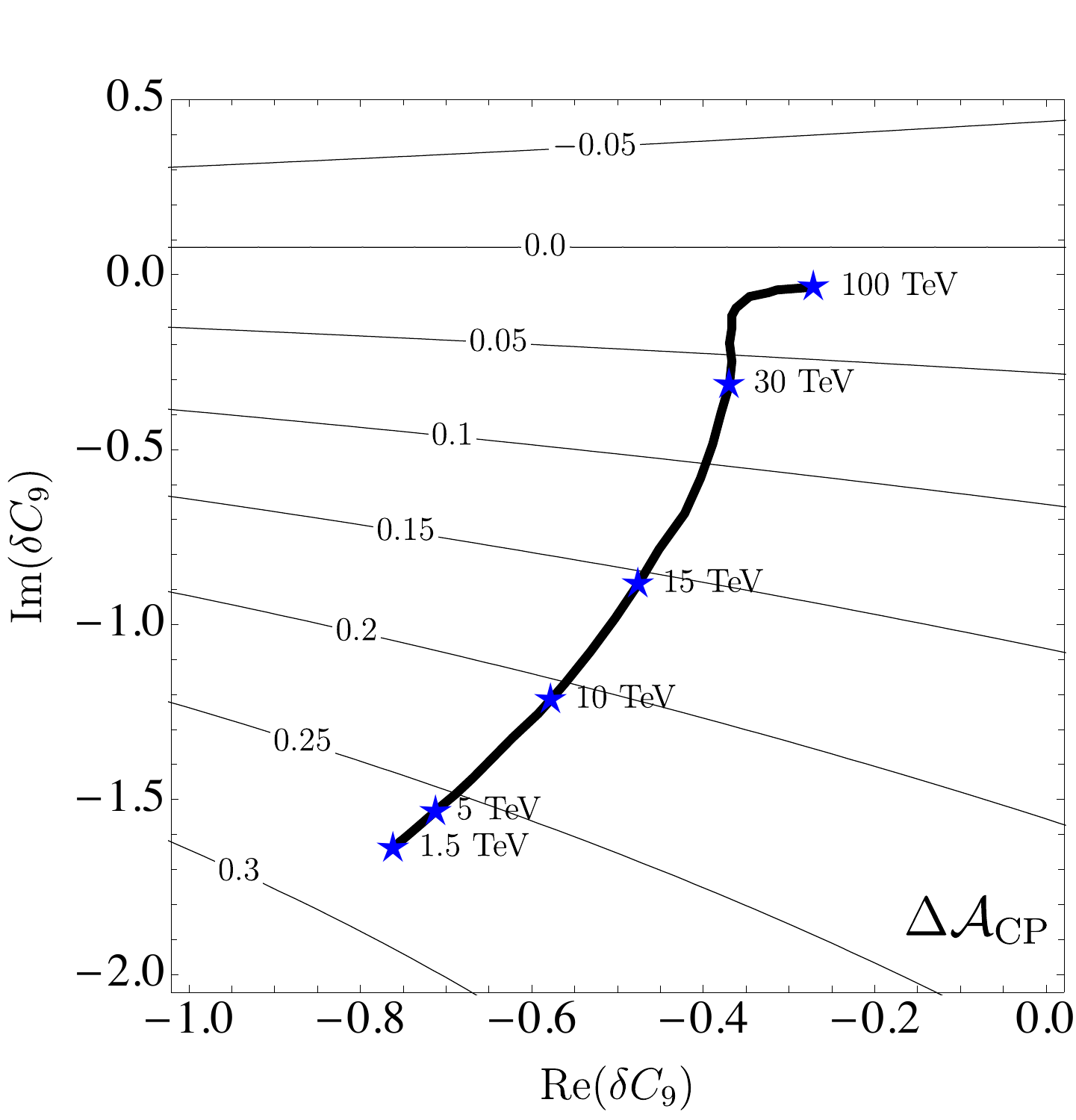}
	\caption{The predictions of the resonantly enhanced CP asymmetry in $B\to K \mu \mu$ (Eq.~\eqref{eq:deltaACP}) throughout the complex $\delta C_9$ plane. The black line shows the positions of the benchmark points from the combined fit (see Fig.~\ref{fig:C9S3}) for $m_{S_3}$ between $1.5~\tev$ and $100~\tev$, with blue stars denoting some of the concrete values of $m_{S_3}$.}
	\label{fig:acppred}
\end{figure}
In Ref.~\cite{Becirevic:2020ssj} it was proposed that direct CP asymmetries $\acp$ in $B\to K \mu \mu$ are enhanced due to the interference effects with narrow charmonium resonances. It was observed that one bin just below ($q^2\in [8,9]~\mathrm{GeV}^2$) and another just above ($q^2\in [10,11]~\mathrm{GeV}^2$) the $J/\psi$ resonance, which have not been previously considered as signal region in experimental
measurements~\cite{1408.0978}, offer a significantly enhanced sensitivity to CP violating NP entering $b\to s \mu \mu$.
The CP asymmetries $\acp^{[8,9]}$ and $\acp^{[10,11]}$ around the $J/\psi$ peak are defined as
\begin{align}
	\label{eq:aCPres}
	\acp^{[q^2_\mrm{min},q^2_\mrm{max}]} \equiv \frac{\bar \Gamma_{[q^2_\mrm{min},q^2_\mrm{max}]}-\Gamma_{[q^2_\mrm{min},q^2_\mrm{max}]}}{\bar \Gamma_{[q^2_\mrm{min},q^2_\mrm{max}]}+\Gamma_{[q^2_\mrm{min},q^2_\mrm{max}]}},
\end{align}
where $\bar \Gamma_{[q^2_\mrm{min},q^2_\mrm{max}]}$ ($\Gamma_{[q^2_\mrm{min},q^2_\mrm{max}]}$) denotes the partial width of the decay $\bar B \to \bar K \mu^+ \mu^-$ (or its CP-conjugated mode), integrated in the kinematical region $q^2_\mrm{min} < q^2 < q^2_\mrm{max}$. We do not write the units $\gev$ in the sub- and superscripts so as not to clutter the notation. As noted in~\cite{Becirevic:2020ssj} the two CP-asymmetries have opposite sign since they are separated by the $J/\psi$ peak, suggesting that we should subtract them in order to further enhance sensitivity to $\im(\delta C_9)$:
\begin{align}
	\label{eq:deltaACP}
  \Delta \acp &\equiv  \frac{\bar \Gamma_{[8,9]} - \Gamma_{[8,9]} - \bar \Gamma_{[10,11]} + \Gamma_{[10,11]}}{\bar \Gamma_{[8,9]} + \Gamma_{[8,9]} + \bar \Gamma_{[10,11]} + \Gamma_{[10,11]}}.\\
  &\nn
      \end{align}
We give simplified numerical formulae in terms of $\delta C_9$ for the above-defined CP asymmetries, where we assume a negative strong phase $\delta_{J/\psi}$ (see Refs.~\cite{Becirevic:2020ssj, 1612.06764})\footnote{The choice of the $\delta_{J/\psi}<0$ solution of the fit in Ref.~\cite{1612.06764} implies predicted $\acp$, compatible with the preferred direction from $A_{7,8,9}$ in Fig.~\ref{fig:C9S3}.}:
\begin{widetext}
	\begin{align}
		\acp^{[8,9]} &= \frac{0.0119(3) -0.153(4)\,\im(\delta C_9)}{1 + 0.407(5)\,\re(\delta C_9)  -0.0084(2)\, \im(\delta C_9) + 0.055(2)\,|\delta C_9|^2},\\
		\acp^{[10,11]} &= \frac{-0.0097(4) + 0.125(5)\,\im(\delta C_9)}{1 + 0.414(5)\,\re(\delta C_9)  -0.0081(3)\, \im(\delta C_9) + 0.053(2)\,|\delta C_9|^2},\\
		\Delta \acp
		&= \frac{0.0108(2) - 0.139(3)\,\im(\delta C_9)}{1 + 0.414(5)\,\re(\delta C_9)  -0.0082(1)\, \im(\delta C_9) + 0.054(1)\,|\delta C_9|^2}.
	\end{align}
      \end{widetext}
The errors in the coefficients are determined by the propagation of uncertainties of the resonant parameters that enter the amplitude~\cite{1612.06764,Becirevic:2020ssj}. In Fig.~\ref{fig:acppred} we show the predictions of $\Delta \acp$ in the $\delta C_9$ complex plane with black contours. Additionally, we show with the bold black line the position of the best fit points with maximal $\im(\delta C_9)$ from considering the combined constraints on Fig.~\ref{fig:C9S3} for $m_{S_3}$ between $1.5~\mathrm{TeV}$ and $100~\mathrm{TeV}$. It is worth noting that the perturbativity constraint on the LQ Yukawa couplings only starts playing a role at $m_{S_3} \gtrsim 200~\tev$.

Furthermore, we consider the observables sensitive to NP scenarios with CPV phases from flavour-tagged angular analysis of $B_d \to K_S \ell \ell$, proposed in Ref.~\cite{2008.08000}, namely $\sigma_0$ and $\sigma_2$. For definitions we refer the reader to Ref.~\cite{2008.08000}, and instead provide a simplified formula for the $\delta C_9 = -\delta C_{10}$ scenario as
\begin{widetext}
	\begin{equation}
		\begin{split}
			\sigma_0 &= \frac{0.37-0.02\, (\im(\delta C_9))^{2}+\im(\delta C_9)\, (-0.04\, \re(\delta C_9)-0.17)+0.02\, (\re(\delta C_9))^{2}+0.19\, \re(\delta C_9)}{1  + 0.51\, \re(\delta C_9) + 0.06\, |\delta C_9|^{2} },
		\end{split}
	\end{equation}
\end{widetext}
wheras $\sigma_2 \approx -\sigma_0$. We provide the SM prediction, as well as some of the benchmark predictions at various masses $m_{S_3}$ of the observable $\sigma_0$ as: $\sigma_0 = \{0.51, 0.49, 0.42, 0.38, 0.37\}$ for $\{1.5~\mathrm{TeV}, 15~\tev, 30~\mathrm{TeV}, 100~\mathrm{TeV},\mathrm{SM}\}$, respectively. The predictions are conservative, as the uncertainties associated with hadronic effects are smaller than the last provided significant digit (Cf.~\cite{2008.08000} for a detailed discussion). 

Lastly, we point out the CP-sensitive observable $A_{\Delta \Gamma_s}^{\mu\mu\gamma}$ in $B_s\to \mu \mu \gamma$ effective lifetime measurements, which could provide additional constraints of CP-violating effects in $b\to s \mu\mu$~\cite{Carvunis:2021jga}.

\section{Conclusion}
\label{sec:conc}
The persistent hints of LFU violation in $b\to s \ell \ell$ may imply an existence of leptoquarks close to the TeV scale that couple to $b \mu$ and $s \mu$. These leptoquark Yukawa couplings, necessary for the explanation of $R_{K^{(\ast)}}$, can in full generality be complex and thus provide a new source of CP violation.

In this Letter we have considered a scalar weak-triplet leptoquark $S_3$ whose left-handed couplings imply the favourable $\delta C_9 = -\delta C_{10}$ scenario. We have shown that large CP violating phases in $S_3$ couplings are bounded neither by $\rkkst$ nor by $\mathcal{B}(B_s \to \mu \mu)$, whereas existing measurements of CP asymmetries $A_{7,8,9}$ constrain $\im(\delta C_9)$, preferably towards negative values. We have considered additional signatures of the leptoquark in the $B_s$ mixing frequency, as well as in the mixing induced CP asymmetry $S_{\psi \phi}$. We have observed that CP-odd effects of the leptoquark in $b\to s\mu\mu$ do not necessarily entail CPV in $B_s$ mixing. For light $S_3$ the $B_s$ mixing does not play an important role, whereas for masses above $5\e{TeV}$ it significantly constrains the imaginary part of $\delta C_9$. For each mass $m_{S_3}$ we have determined the maximal allowed CPV coupling $\im(\delta C_9)$ and derived corresponding predictions. We have pointed out potential measurements which could help pinpoint CPV in $b\to s \ell \ell$ in the future: enhanced CP asymmetries in $B\to K \mu \mu$ in the vicinity of the $J/\psi$ resonance~\cite{Becirevic:2020ssj}, and flavour-tagged measurements of $B_d\to K_S \ell \ell$~\cite{2008.08000}. Such measurements, as well as increased theoretical precision in the $B_s$ mixing observables, could help determine whether the hints of LFU violation also hint at new sources of CP violation in the Universe.

\section*{Acknowledgments}
The project was in part financially supported by the Slovenian Research Agency (research core funding No.~P1-0035). This work is supported by the COST Action CA16201 PARTICLEFACE (European Cooperation in Science and Technology). A.~S. is supported by the Young Researchers Programme of the Slovenian Research Agency under the grant No.~50510, core funding grant P1-0035.

\bibliography{S3_CPV}{}

\begin{thebibliography}{51}
\expandafter\ifx\csname natexlab\endcsname\relax\def\natexlab#1{#1}\fi
\expandafter\ifx\csname bibnamefont\endcsname\relax
  \def\bibnamefont#1{#1}\fi
\expandafter\ifx\csname bibfnamefont\endcsname\relax
  \def\bibfnamefont#1{#1}\fi
\expandafter\ifx\csname citenamefont\endcsname\relax
  \def\citenamefont#1{#1}\fi
\expandafter\ifx\csname url\endcsname\relax
  \def\url#1{\texttt{#1}}\fi
\expandafter\ifx\csname urlprefix\endcsname\relax\def\urlprefix{URL }\fi
\providecommand{\bibinfo}[2]{#2}
\providecommand{\eprint}[2][]{\url{#2}}

\bibitem[{\citenamefont{Hiller and Kruger}(2004)}]{hep-ph/0310219}
\bibinfo{author}{\bibfnamefont{G.}~\bibnamefont{Hiller}} \bibnamefont{and}
  \bibinfo{author}{\bibfnamefont{F.}~\bibnamefont{Kruger}},
  \bibinfo{journal}{Phys. Rev. D} \textbf{\bibinfo{volume}{69}},
  \bibinfo{pages}{074020} (\bibinfo{year}{2004}), \eprint{hep-ph/0310219}.

\bibitem[{\citenamefont{Hiller and Schmaltz}(2014)}]{1408.1627}
\bibinfo{author}{\bibfnamefont{G.}~\bibnamefont{Hiller}} \bibnamefont{and}
  \bibinfo{author}{\bibfnamefont{M.}~\bibnamefont{Schmaltz}},
  \bibinfo{journal}{Phys. Rev. D} \textbf{\bibinfo{volume}{90}},
  \bibinfo{pages}{054014} (\bibinfo{year}{2014}), \eprint{1408.1627}.

\bibitem[{\citenamefont{Bifani et~al.}(2019)\citenamefont{Bifani,
  Descotes-Genon, Romero~Vidal, and Schune}}]{1809.06229}
\bibinfo{author}{\bibfnamefont{S.}~\bibnamefont{Bifani}},
  \bibinfo{author}{\bibfnamefont{S.}~\bibnamefont{Descotes-Genon}},
  \bibinfo{author}{\bibfnamefont{A.}~\bibnamefont{Romero~Vidal}},
  \bibnamefont{and} \bibinfo{author}{\bibfnamefont{M.-H.}
  \bibnamefont{Schune}}, \bibinfo{journal}{J. Phys. G}
  \textbf{\bibinfo{volume}{46}}, \bibinfo{pages}{023001}
  (\bibinfo{year}{2019}), \eprint{1809.06229}.

\bibitem[{\citenamefont{Sirunyan et~al.}(2021)}]{CMS:2021ctt}
\bibinfo{author}{\bibfnamefont{A.~M.} \bibnamefont{Sirunyan}}
  \bibnamefont{et~al.} (\bibinfo{collaboration}{CMS}), \bibinfo{journal}{JHEP}
  \textbf{\bibinfo{volume}{07}}, \bibinfo{pages}{208} (\bibinfo{year}{2021}),
  \eprint{2103.02708}.

\bibitem[{\citenamefont{Aaij et~al.}(2014{\natexlab{a}})}]{Aaij:2014ora}
\bibinfo{author}{\bibfnamefont{R.}~\bibnamefont{Aaij}} \bibnamefont{et~al.}
  (\bibinfo{collaboration}{LHCb}), \bibinfo{journal}{Phys. Rev. Lett.}
  \textbf{\bibinfo{volume}{113}}, \bibinfo{pages}{151601}
  (\bibinfo{year}{2014}{\natexlab{a}}), \eprint{1406.6482}.

\bibitem[{\citenamefont{Aaij et~al.}(2019)}]{Aaij:2019wad}
\bibinfo{author}{\bibfnamefont{R.}~\bibnamefont{Aaij}} \bibnamefont{et~al.}
  (\bibinfo{collaboration}{LHCb}), \bibinfo{journal}{Phys. Rev. Lett.}
  \textbf{\bibinfo{volume}{122}}, \bibinfo{pages}{191801}
  (\bibinfo{year}{2019}), \eprint{1903.09252}.

\bibitem[{\citenamefont{Lancierini}(2021)}]{Lancierini:2021aek}
\bibinfo{author}{\bibfnamefont{D.}~\bibnamefont{Lancierini}}
  (\bibinfo{year}{2021}), \eprint{2105.10303}.

\bibitem[{\citenamefont{Aaij et~al.}(2017{\natexlab{a}})}]{Aaij:2017vbb}
\bibinfo{author}{\bibfnamefont{R.}~\bibnamefont{Aaij}} \bibnamefont{et~al.}
  (\bibinfo{collaboration}{LHCb}), \bibinfo{journal}{JHEP}
  \textbf{\bibinfo{volume}{08}}, \bibinfo{pages}{055}
  (\bibinfo{year}{2017}{\natexlab{a}}), \eprint{1705.05802}.

\bibitem[{\citenamefont{Bordone et~al.}(2016)\citenamefont{Bordone, Isidori,
  and Pattori}}]{Bordone:2016gaq}
\bibinfo{author}{\bibfnamefont{M.}~\bibnamefont{Bordone}},
  \bibinfo{author}{\bibfnamefont{G.}~\bibnamefont{Isidori}}, \bibnamefont{and}
  \bibinfo{author}{\bibfnamefont{A.}~\bibnamefont{Pattori}},
  \bibinfo{journal}{Eur. Phys. J. C} \textbf{\bibinfo{volume}{76}},
  \bibinfo{pages}{440} (\bibinfo{year}{2016}), \eprint{1605.07633}.

\bibitem[{\citenamefont{Hurth et~al.}(2021)\citenamefont{Hurth, Mahmoudi,
  Santos, and Neshatpour}}]{2104.10058}
\bibinfo{author}{\bibfnamefont{T.}~\bibnamefont{Hurth}},
  \bibinfo{author}{\bibfnamefont{F.}~\bibnamefont{Mahmoudi}},
  \bibinfo{author}{\bibfnamefont{D.~M.} \bibnamefont{Santos}},
  \bibnamefont{and}
  \bibinfo{author}{\bibfnamefont{S.}~\bibnamefont{Neshatpour}}
  (\bibinfo{year}{2021}), \eprint{2104.10058}.

\bibitem[{\citenamefont{Altmannshofer and
  Stangl}(2021)}]{Altmannshofer:2021qrr}
\bibinfo{author}{\bibfnamefont{W.}~\bibnamefont{Altmannshofer}}
  \bibnamefont{and} \bibinfo{author}{\bibfnamefont{P.}~\bibnamefont{Stangl}}
  (\bibinfo{year}{2021}), \eprint{2103.13370}.

\bibitem[{\citenamefont{Angelescu
  et~al.}(2021{\natexlab{a}})\citenamefont{Angelescu, Be\v{c}irevi\'c,
  Faroughy, Jaffredo, and Sumensari}}]{Angelescu:2021lln}
\bibinfo{author}{\bibfnamefont{A.}~\bibnamefont{Angelescu}},
  \bibinfo{author}{\bibfnamefont{D.}~\bibnamefont{Be\v{c}irevi\'c}},
  \bibinfo{author}{\bibfnamefont{D.~A.} \bibnamefont{Faroughy}},
  \bibinfo{author}{\bibfnamefont{F.}~\bibnamefont{Jaffredo}}, \bibnamefont{and}
  \bibinfo{author}{\bibfnamefont{O.}~\bibnamefont{Sumensari}}
  (\bibinfo{year}{2021}{\natexlab{a}}), \eprint{2103.12504}.

\bibitem[{\citenamefont{Carvunis et~al.}(2021)\citenamefont{Carvunis, Dettori,
  Gangal, Guadagnoli, and Normand}}]{Carvunis:2021jga}
\bibinfo{author}{\bibfnamefont{A.}~\bibnamefont{Carvunis}},
  \bibinfo{author}{\bibfnamefont{F.}~\bibnamefont{Dettori}},
  \bibinfo{author}{\bibfnamefont{S.}~\bibnamefont{Gangal}},
  \bibinfo{author}{\bibfnamefont{D.}~\bibnamefont{Guadagnoli}},
  \bibnamefont{and} \bibinfo{author}{\bibfnamefont{C.}~\bibnamefont{Normand}}
  (\bibinfo{year}{2021}), \eprint{2102.13390}.

\bibitem[{\citenamefont{Be\v{c}irevi\'c
  et~al.}(2020)\citenamefont{Be\v{c}irevi\'c, Fajfer, Ko\v{s}nik, and
  Smolkovi\v{c}}}]{Becirevic:2020ssj}
\bibinfo{author}{\bibfnamefont{D.}~\bibnamefont{Be\v{c}irevi\'c}},
  \bibinfo{author}{\bibfnamefont{S.}~\bibnamefont{Fajfer}},
  \bibinfo{author}{\bibfnamefont{N.}~\bibnamefont{Ko\v{s}nik}},
  \bibnamefont{and}
  \bibinfo{author}{\bibfnamefont{A.}~\bibnamefont{Smolkovi\v{c}}},
  \bibinfo{journal}{Eur. Phys. J. C} \textbf{\bibinfo{volume}{80}},
  \bibinfo{pages}{940} (\bibinfo{year}{2020}), \eprint{2008.09064}.

\bibitem[{\citenamefont{Alok et~al.}(2017)\citenamefont{Alok, Bhattacharya,
  Kumar, Kumar, London, and Sankar}}]{1703.09247}
\bibinfo{author}{\bibfnamefont{A.~K.} \bibnamefont{Alok}},
  \bibinfo{author}{\bibfnamefont{B.}~\bibnamefont{Bhattacharya}},
  \bibinfo{author}{\bibfnamefont{D.}~\bibnamefont{Kumar}},
  \bibinfo{author}{\bibfnamefont{J.}~\bibnamefont{Kumar}},
  \bibinfo{author}{\bibfnamefont{D.}~\bibnamefont{London}}, \bibnamefont{and}
  \bibinfo{author}{\bibfnamefont{S.~U.} \bibnamefont{Sankar}},
  \bibinfo{journal}{Phys. Rev. D} \textbf{\bibinfo{volume}{96}},
  \bibinfo{pages}{015034} (\bibinfo{year}{2017}), \eprint{1703.09247}.

\bibitem[{\citenamefont{Di~Luzio et~al.}(2018)\citenamefont{Di~Luzio, Kirk, and
  Lenz}}]{1712.06572}
\bibinfo{author}{\bibfnamefont{L.}~\bibnamefont{Di~Luzio}},
  \bibinfo{author}{\bibfnamefont{M.}~\bibnamefont{Kirk}}, \bibnamefont{and}
  \bibinfo{author}{\bibfnamefont{A.}~\bibnamefont{Lenz}},
  \bibinfo{journal}{Phys. Rev. D} \textbf{\bibinfo{volume}{97}},
  \bibinfo{pages}{095035} (\bibinfo{year}{2018}), \eprint{1712.06572}.

\bibitem[{\citenamefont{Di~Luzio et~al.}(2019)\citenamefont{Di~Luzio, Kirk,
  Lenz, and Rauh}}]{1909.11087}
\bibinfo{author}{\bibfnamefont{L.}~\bibnamefont{Di~Luzio}},
  \bibinfo{author}{\bibfnamefont{M.}~\bibnamefont{Kirk}},
  \bibinfo{author}{\bibfnamefont{A.}~\bibnamefont{Lenz}}, \bibnamefont{and}
  \bibinfo{author}{\bibfnamefont{T.}~\bibnamefont{Rauh}},
  \bibinfo{journal}{JHEP} \textbf{\bibinfo{volume}{12}}, \bibinfo{pages}{009}
  (\bibinfo{year}{2019}), \eprint{1909.11087}.

\bibitem[{\citenamefont{Crivellin et~al.}(2017)\citenamefont{Crivellin,
  M\"uller, and Ota}}]{1703.09226}
\bibinfo{author}{\bibfnamefont{A.}~\bibnamefont{Crivellin}},
  \bibinfo{author}{\bibfnamefont{D.}~\bibnamefont{M\"uller}}, \bibnamefont{and}
  \bibinfo{author}{\bibfnamefont{T.}~\bibnamefont{Ota}},
  \bibinfo{journal}{JHEP} \textbf{\bibinfo{volume}{09}}, \bibinfo{pages}{040}
  (\bibinfo{year}{2017}), \eprint{1703.09226}.

\bibitem[{\citenamefont{Be\v{c}irevi\'c
  et~al.}(2018)\citenamefont{Be\v{c}irevi\'c, Dor\v{s}ner, Fajfer, Ko\v{s}nik,
  Faroughy, and Sumensari}}]{Becirevic:2018afm}
\bibinfo{author}{\bibfnamefont{D.}~\bibnamefont{Be\v{c}irevi\'c}},
  \bibinfo{author}{\bibfnamefont{I.}~\bibnamefont{Dor\v{s}ner}},
  \bibinfo{author}{\bibfnamefont{S.}~\bibnamefont{Fajfer}},
  \bibinfo{author}{\bibfnamefont{N.}~\bibnamefont{Ko\v{s}nik}},
  \bibinfo{author}{\bibfnamefont{D.~A.} \bibnamefont{Faroughy}},
  \bibnamefont{and}
  \bibinfo{author}{\bibfnamefont{O.}~\bibnamefont{Sumensari}},
  \bibinfo{journal}{Phys. Rev. D} \textbf{\bibinfo{volume}{98}},
  \bibinfo{pages}{055003} (\bibinfo{year}{2018}), \eprint{1806.05689}.

\bibitem[{\citenamefont{Crivellin et~al.}(2020)\citenamefont{Crivellin,
  M\"uller, and Saturnino}}]{Crivellin:2019dwb}
\bibinfo{author}{\bibfnamefont{A.}~\bibnamefont{Crivellin}},
  \bibinfo{author}{\bibfnamefont{D.}~\bibnamefont{M\"uller}}, \bibnamefont{and}
  \bibinfo{author}{\bibfnamefont{F.}~\bibnamefont{Saturnino}},
  \bibinfo{journal}{JHEP} \textbf{\bibinfo{volume}{06}}, \bibinfo{pages}{020}
  (\bibinfo{year}{2020}), \eprint{1912.04224}.

\bibitem[{\citenamefont{Gherardi et~al.}(2021)\citenamefont{Gherardi, Marzocca,
  and Venturini}}]{2008.09548}
\bibinfo{author}{\bibfnamefont{V.}~\bibnamefont{Gherardi}},
  \bibinfo{author}{\bibfnamefont{D.}~\bibnamefont{Marzocca}}, \bibnamefont{and}
  \bibinfo{author}{\bibfnamefont{E.}~\bibnamefont{Venturini}},
  \bibinfo{journal}{JHEP} \textbf{\bibinfo{volume}{01}}, \bibinfo{pages}{138}
  (\bibinfo{year}{2021}), \eprint{2008.09548}.

\bibitem[{\citenamefont{Angelescu
  et~al.}(2021{\natexlab{b}})\citenamefont{Angelescu, Be\v{c}irevi\'c,
  Faroughy, Jaffredo, and Sumensari}}]{2103.12504}
\bibinfo{author}{\bibfnamefont{A.}~\bibnamefont{Angelescu}},
  \bibinfo{author}{\bibfnamefont{D.}~\bibnamefont{Be\v{c}irevi\'c}},
  \bibinfo{author}{\bibfnamefont{D.~A.} \bibnamefont{Faroughy}},
  \bibinfo{author}{\bibfnamefont{F.}~\bibnamefont{Jaffredo}}, \bibnamefont{and}
  \bibinfo{author}{\bibfnamefont{O.}~\bibnamefont{Sumensari}}
  (\bibinfo{year}{2021}{\natexlab{b}}), \eprint{2103.12504}.

\bibitem[{\citenamefont{Choi et~al.}(2018)\citenamefont{Choi, Kang, Lee, and
  Ro}}]{Choi:2018stw}
\bibinfo{author}{\bibfnamefont{S.-M.} \bibnamefont{Choi}},
  \bibinfo{author}{\bibfnamefont{Y.-J.} \bibnamefont{Kang}},
  \bibinfo{author}{\bibfnamefont{H.~M.} \bibnamefont{Lee}}, \bibnamefont{and}
  \bibinfo{author}{\bibfnamefont{T.-G.} \bibnamefont{Ro}},
  \bibinfo{journal}{JHEP} \textbf{\bibinfo{volume}{10}}, \bibinfo{pages}{104}
  (\bibinfo{year}{2018}), \eprint{1807.06547}.

\bibitem[{\citenamefont{D'Eramo et~al.}(2021)\citenamefont{D'Eramo, Ko\v{s}nik,
  Pobbe, Smolkovi\v{c}, and Sumensari}}]{DEramo:2020sqv}
\bibinfo{author}{\bibfnamefont{F.}~\bibnamefont{D'Eramo}},
  \bibinfo{author}{\bibfnamefont{N.}~\bibnamefont{Ko\v{s}nik}},
  \bibinfo{author}{\bibfnamefont{F.}~\bibnamefont{Pobbe}},
  \bibinfo{author}{\bibfnamefont{A.}~\bibnamefont{Smolkovi\v{c}}},
  \bibnamefont{and}
  \bibinfo{author}{\bibfnamefont{O.}~\bibnamefont{Sumensari}},
  \bibinfo{journal}{Phys. Rev. D} \textbf{\bibinfo{volume}{104}},
  \bibinfo{pages}{015035} (\bibinfo{year}{2021}), \eprint{2012.05743}.

\bibitem[{\citenamefont{Grzadkowski et~al.}(2010)\citenamefont{Grzadkowski,
  Iskrzynski, Misiak, and Rosiek}}]{Grzadkowski:2010es}
\bibinfo{author}{\bibfnamefont{B.}~\bibnamefont{Grzadkowski}},
  \bibinfo{author}{\bibfnamefont{M.}~\bibnamefont{Iskrzynski}},
  \bibinfo{author}{\bibfnamefont{M.}~\bibnamefont{Misiak}}, \bibnamefont{and}
  \bibinfo{author}{\bibfnamefont{J.}~\bibnamefont{Rosiek}},
  \bibinfo{journal}{JHEP} \textbf{\bibinfo{volume}{10}}, \bibinfo{pages}{085}
  (\bibinfo{year}{2010}), \eprint{1008.4884}.

\bibitem[{\citenamefont{Jenkins et~al.}(2013)\citenamefont{Jenkins, Manohar,
  and Trott}}]{1308.2627}
\bibinfo{author}{\bibfnamefont{E.~E.} \bibnamefont{Jenkins}},
  \bibinfo{author}{\bibfnamefont{A.~V.} \bibnamefont{Manohar}},
  \bibnamefont{and} \bibinfo{author}{\bibfnamefont{M.}~\bibnamefont{Trott}},
  \bibinfo{journal}{JHEP} \textbf{\bibinfo{volume}{10}}, \bibinfo{pages}{087}
  (\bibinfo{year}{2013}), \eprint{1308.2627}.

\bibitem[{\citenamefont{Jenkins et~al.}(2014)\citenamefont{Jenkins, Manohar,
  and Trott}}]{1310.4838}
\bibinfo{author}{\bibfnamefont{E.~E.} \bibnamefont{Jenkins}},
  \bibinfo{author}{\bibfnamefont{A.~V.} \bibnamefont{Manohar}},
  \bibnamefont{and} \bibinfo{author}{\bibfnamefont{M.}~\bibnamefont{Trott}},
  \bibinfo{journal}{JHEP} \textbf{\bibinfo{volume}{01}}, \bibinfo{pages}{035}
  (\bibinfo{year}{2014}), \eprint{1310.4838}.

\bibitem[{\citenamefont{Alonso et~al.}(2014)\citenamefont{Alonso, Jenkins,
  Manohar, and Trott}}]{1312.2014}
\bibinfo{author}{\bibfnamefont{R.}~\bibnamefont{Alonso}},
  \bibinfo{author}{\bibfnamefont{E.~E.} \bibnamefont{Jenkins}},
  \bibinfo{author}{\bibfnamefont{A.~V.} \bibnamefont{Manohar}},
  \bibnamefont{and} \bibinfo{author}{\bibfnamefont{M.}~\bibnamefont{Trott}},
  \bibinfo{journal}{JHEP} \textbf{\bibinfo{volume}{04}}, \bibinfo{pages}{159}
  (\bibinfo{year}{2014}), \eprint{1312.2014}.

\bibitem[{\citenamefont{Dor\v{s}ner et~al.}(2016)\citenamefont{Dor\v{s}ner,
  Fajfer, Greljo, Kamenik, and Ko\v{s}nik}}]{1603.04993}
\bibinfo{author}{\bibfnamefont{I.}~\bibnamefont{Dor\v{s}ner}},
  \bibinfo{author}{\bibfnamefont{S.}~\bibnamefont{Fajfer}},
  \bibinfo{author}{\bibfnamefont{A.}~\bibnamefont{Greljo}},
  \bibinfo{author}{\bibfnamefont{J.~F.} \bibnamefont{Kamenik}},
  \bibnamefont{and}
  \bibinfo{author}{\bibfnamefont{N.}~\bibnamefont{Ko\v{s}nik}},
  \bibinfo{journal}{Phys. Rept.} \textbf{\bibinfo{volume}{641}},
  \bibinfo{pages}{1} (\bibinfo{year}{2016}), \eprint{1603.04993}.

\bibitem[{\citenamefont{Gherardi et~al.}(2020)\citenamefont{Gherardi, Marzocca,
  and Venturini}}]{2003.12525}
\bibinfo{author}{\bibfnamefont{V.}~\bibnamefont{Gherardi}},
  \bibinfo{author}{\bibfnamefont{D.}~\bibnamefont{Marzocca}}, \bibnamefont{and}
  \bibinfo{author}{\bibfnamefont{E.}~\bibnamefont{Venturini}},
  \bibinfo{journal}{JHEP} \textbf{\bibinfo{volume}{07}}, \bibinfo{pages}{225}
  (\bibinfo{year}{2020}), \bibinfo{note}{[Erratum: JHEP 01, 006 (2021)]},
  \eprint{2003.12525}.

\bibitem[{\citenamefont{Buras et~al.}(1994)\citenamefont{Buras, Misiak, Munz,
  and Pokorski}}]{hep-ph/9311345}
\bibinfo{author}{\bibfnamefont{A.~J.} \bibnamefont{Buras}},
  \bibinfo{author}{\bibfnamefont{M.}~\bibnamefont{Misiak}},
  \bibinfo{author}{\bibfnamefont{M.}~\bibnamefont{Munz}}, \bibnamefont{and}
  \bibinfo{author}{\bibfnamefont{S.}~\bibnamefont{Pokorski}},
  \bibinfo{journal}{Nucl. Phys. B} \textbf{\bibinfo{volume}{424}},
  \bibinfo{pages}{374} (\bibinfo{year}{1994}), \eprint{hep-ph/9311345}.

\bibitem[{\citenamefont{Altmannshofer et~al.}(2009)\citenamefont{Altmannshofer,
  Ball, Bharucha, Buras, Straub, and Wick}}]{0811.1214}
\bibinfo{author}{\bibfnamefont{W.}~\bibnamefont{Altmannshofer}},
  \bibinfo{author}{\bibfnamefont{P.}~\bibnamefont{Ball}},
  \bibinfo{author}{\bibfnamefont{A.}~\bibnamefont{Bharucha}},
  \bibinfo{author}{\bibfnamefont{A.~J.} \bibnamefont{Buras}},
  \bibinfo{author}{\bibfnamefont{D.~M.} \bibnamefont{Straub}},
  \bibnamefont{and} \bibinfo{author}{\bibfnamefont{M.}~\bibnamefont{Wick}},
  \bibinfo{journal}{JHEP} \textbf{\bibinfo{volume}{01}}, \bibinfo{pages}{019}
  (\bibinfo{year}{2009}), \eprint{0811.1214}.

\bibitem[{\citenamefont{Gonz\'alez-Alonso
  et~al.}(2017)\citenamefont{Gonz\'alez-Alonso, Martin~Camalich, and
  Mimouni}}]{1706.00410}
\bibinfo{author}{\bibfnamefont{M.}~\bibnamefont{Gonz\'alez-Alonso}},
  \bibinfo{author}{\bibfnamefont{J.}~\bibnamefont{Martin~Camalich}},
  \bibnamefont{and} \bibinfo{author}{\bibfnamefont{K.}~\bibnamefont{Mimouni}},
  \bibinfo{journal}{Phys. Lett. B} \textbf{\bibinfo{volume}{772}},
  \bibinfo{pages}{777} (\bibinfo{year}{2017}), \eprint{1706.00410}.

\bibitem[{\citenamefont{Artuso et~al.}(2016)\citenamefont{Artuso, Borissov, and
  Lenz}}]{1511.09466}
\bibinfo{author}{\bibfnamefont{M.}~\bibnamefont{Artuso}},
  \bibinfo{author}{\bibfnamefont{G.}~\bibnamefont{Borissov}}, \bibnamefont{and}
  \bibinfo{author}{\bibfnamefont{A.}~\bibnamefont{Lenz}},
  \bibinfo{journal}{Rev. Mod. Phys.} \textbf{\bibinfo{volume}{88}},
  \bibinfo{pages}{045002} (\bibinfo{year}{2016}), \bibinfo{note}{[Addendum:
  Rev.Mod.Phys. 91, 049901 (2019)]}, \eprint{1511.09466}.

\bibitem[{\citenamefont{Buras et~al.}(1990)\citenamefont{Buras, Jamin, and
  Weisz}}]{Buras:1990fn}
\bibinfo{author}{\bibfnamefont{A.~J.} \bibnamefont{Buras}},
  \bibinfo{author}{\bibfnamefont{M.}~\bibnamefont{Jamin}}, \bibnamefont{and}
  \bibinfo{author}{\bibfnamefont{P.~H.} \bibnamefont{Weisz}},
  \bibinfo{journal}{Nucl. Phys. B} \textbf{\bibinfo{volume}{347}},
  \bibinfo{pages}{491} (\bibinfo{year}{1990}).

\bibitem[{\citenamefont{Aaij et~al.}(2021)}]{LHCb:2021awg}
\bibinfo{author}{\bibfnamefont{R.}~\bibnamefont{Aaij}} \bibnamefont{et~al.}
  (\bibinfo{collaboration}{LHCb}) (\bibinfo{year}{2021}), \eprint{2108.09283}.

\bibitem[{\citenamefont{Aaboud et~al.}(2019)}]{1812.03017}
\bibinfo{author}{\bibfnamefont{M.}~\bibnamefont{Aaboud}} \bibnamefont{et~al.}
  (\bibinfo{collaboration}{ATLAS}), \bibinfo{journal}{JHEP}
  \textbf{\bibinfo{volume}{04}}, \bibinfo{pages}{098} (\bibinfo{year}{2019}),
  \eprint{1812.03017}.

\bibitem[{\citenamefont{Sirunyan et~al.}(2020)}]{1910.12127}
\bibinfo{author}{\bibfnamefont{A.~M.} \bibnamefont{Sirunyan}}
  \bibnamefont{et~al.} (\bibinfo{collaboration}{CMS}), \bibinfo{journal}{JHEP}
  \textbf{\bibinfo{volume}{04}}, \bibinfo{pages}{188} (\bibinfo{year}{2020}),
  \eprint{1910.12127}.

\bibitem[{\citenamefont{Alguer\'o et~al.}(2021)\citenamefont{Alguer\'o,
  Capdevila, Descotes-Genon, Matias, and Novoa-Brunet}}]{2104.08921}
\bibinfo{author}{\bibfnamefont{M.}~\bibnamefont{Alguer\'o}},
  \bibinfo{author}{\bibfnamefont{B.}~\bibnamefont{Capdevila}},
  \bibinfo{author}{\bibfnamefont{S.}~\bibnamefont{Descotes-Genon}},
  \bibinfo{author}{\bibfnamefont{J.}~\bibnamefont{Matias}}, \bibnamefont{and}
  \bibinfo{author}{\bibfnamefont{M.}~\bibnamefont{Novoa-Brunet}}, in
  \emph{\bibinfo{booktitle}{{55th Rencontres de Moriond on QCD and High Energy
  Interactions}}} (\bibinfo{year}{2021}), \eprint{2104.08921}.

\bibitem[{\citenamefont{Aoki et~al.}(2020)}]{Aoki:2019cca}
\bibinfo{author}{\bibfnamefont{S.}~\bibnamefont{Aoki}} \bibnamefont{et~al.}
  (\bibinfo{collaboration}{Flavour Lattice Averaging Group}),
  \bibinfo{journal}{Eur. Phys. J. C} \textbf{\bibinfo{volume}{80}},
  \bibinfo{pages}{113} (\bibinfo{year}{2020}), \eprint{1902.08191}.

\bibitem[{\citenamefont{De~Bruyn et~al.}(2012)\citenamefont{De~Bruyn,
  Fleischer, Knegjens, Koppenburg, Merk, and Tuning}}]{1204.1735}
\bibinfo{author}{\bibfnamefont{K.}~\bibnamefont{De~Bruyn}},
  \bibinfo{author}{\bibfnamefont{R.}~\bibnamefont{Fleischer}},
  \bibinfo{author}{\bibfnamefont{R.}~\bibnamefont{Knegjens}},
  \bibinfo{author}{\bibfnamefont{P.}~\bibnamefont{Koppenburg}},
  \bibinfo{author}{\bibfnamefont{M.}~\bibnamefont{Merk}}, \bibnamefont{and}
  \bibinfo{author}{\bibfnamefont{N.}~\bibnamefont{Tuning}},
  \bibinfo{journal}{Phys. Rev. D} \textbf{\bibinfo{volume}{86}},
  \bibinfo{pages}{014027} (\bibinfo{year}{2012}), \eprint{1204.1735}.

\bibitem[{\citenamefont{Beneke et~al.}(2019)\citenamefont{Beneke, Bobeth, and
  Szafron}}]{1908.07011}
\bibinfo{author}{\bibfnamefont{M.}~\bibnamefont{Beneke}},
  \bibinfo{author}{\bibfnamefont{C.}~\bibnamefont{Bobeth}}, \bibnamefont{and}
  \bibinfo{author}{\bibfnamefont{R.}~\bibnamefont{Szafron}},
  \bibinfo{journal}{JHEP} \textbf{\bibinfo{volume}{10}}, \bibinfo{pages}{232}
  (\bibinfo{year}{2019}), \eprint{1908.07011}.

\bibitem[{\citenamefont{Bobeth et~al.}(2008)\citenamefont{Bobeth, Hiller, and
  Piranishvili}}]{0805.2525}
\bibinfo{author}{\bibfnamefont{C.}~\bibnamefont{Bobeth}},
  \bibinfo{author}{\bibfnamefont{G.}~\bibnamefont{Hiller}}, \bibnamefont{and}
  \bibinfo{author}{\bibfnamefont{G.}~\bibnamefont{Piranishvili}},
  \bibinfo{journal}{JHEP} \textbf{\bibinfo{volume}{07}}, \bibinfo{pages}{106}
  (\bibinfo{year}{2008}), \eprint{0805.2525}.

\bibitem[{\citenamefont{Aaij et~al.}(2016)}]{1512.04442}
\bibinfo{author}{\bibfnamefont{R.}~\bibnamefont{Aaij}} \bibnamefont{et~al.}
  (\bibinfo{collaboration}{LHCb}), \bibinfo{journal}{JHEP}
  \textbf{\bibinfo{volume}{02}}, \bibinfo{pages}{104} (\bibinfo{year}{2016}),
  \eprint{1512.04442}.

\bibitem[{\citenamefont{Straub}(2018)}]{1810.08132}
\bibinfo{author}{\bibfnamefont{D.~M.} \bibnamefont{Straub}}
  (\bibinfo{year}{2018}), \eprint{1810.08132}.

\bibitem[{\citenamefont{Charles et~al.}(2005)\citenamefont{Charles, Hocker,
  Lacker, Laplace, Le~Diberder, Malcles, Ocariz, Pivk, and
  Roos}}]{hep-ph/0406184}
\bibinfo{author}{\bibfnamefont{J.}~\bibnamefont{Charles}},
  \bibinfo{author}{\bibfnamefont{A.}~\bibnamefont{Hocker}},
  \bibinfo{author}{\bibfnamefont{H.}~\bibnamefont{Lacker}},
  \bibinfo{author}{\bibfnamefont{S.}~\bibnamefont{Laplace}},
  \bibinfo{author}{\bibfnamefont{F.~R.} \bibnamefont{Le~Diberder}},
  \bibinfo{author}{\bibfnamefont{J.}~\bibnamefont{Malcles}},
  \bibinfo{author}{\bibfnamefont{J.}~\bibnamefont{Ocariz}},
  \bibinfo{author}{\bibfnamefont{M.}~\bibnamefont{Pivk}}, \bibnamefont{and}
  \bibinfo{author}{\bibfnamefont{L.}~\bibnamefont{Roos}}
  (\bibinfo{collaboration}{CKMfitter Group}), \bibinfo{journal}{Eur. Phys. J.
  C} \textbf{\bibinfo{volume}{41}}, \bibinfo{pages}{1} (\bibinfo{year}{2005}),
  \eprint{hep-ph/0406184}.

\bibitem[{\citenamefont{Charles et~al.}(2020)\citenamefont{Charles,
  Descotes-Genon, Ligeti, Monteil, Papucci, Trabelsi, and
  Vale~Silva}}]{2006.04824}
\bibinfo{author}{\bibfnamefont{J.}~\bibnamefont{Charles}},
  \bibinfo{author}{\bibfnamefont{S.}~\bibnamefont{Descotes-Genon}},
  \bibinfo{author}{\bibfnamefont{Z.}~\bibnamefont{Ligeti}},
  \bibinfo{author}{\bibfnamefont{S.}~\bibnamefont{Monteil}},
  \bibinfo{author}{\bibfnamefont{M.}~\bibnamefont{Papucci}},
  \bibinfo{author}{\bibfnamefont{K.}~\bibnamefont{Trabelsi}}, \bibnamefont{and}
  \bibinfo{author}{\bibfnamefont{L.}~\bibnamefont{Vale~Silva}},
  \bibinfo{journal}{Phys. Rev. D} \textbf{\bibinfo{volume}{102}},
  \bibinfo{pages}{056023} (\bibinfo{year}{2020}), \eprint{2006.04824}.

\bibitem[{\citenamefont{Amhis et~al.}(2021)}]{HFLAV:2019otj}
\bibinfo{author}{\bibfnamefont{Y.~S.} \bibnamefont{Amhis}} \bibnamefont{et~al.}
  (\bibinfo{collaboration}{HFLAV}), \bibinfo{journal}{Eur. Phys. J. C}
  \textbf{\bibinfo{volume}{81}}, \bibinfo{pages}{226} (\bibinfo{year}{2021}),
  \eprint{1909.12524}.

\bibitem[{\citenamefont{Aaij et~al.}(2014{\natexlab{b}})}]{1408.0978}
\bibinfo{author}{\bibfnamefont{R.}~\bibnamefont{Aaij}} \bibnamefont{et~al.}
  (\bibinfo{collaboration}{LHCb}), \bibinfo{journal}{JHEP}
  \textbf{\bibinfo{volume}{09}}, \bibinfo{pages}{177}
  (\bibinfo{year}{2014}{\natexlab{b}}), \eprint{1408.0978}.

\bibitem[{\citenamefont{Aaij et~al.}(2017{\natexlab{b}})}]{1612.06764}
\bibinfo{author}{\bibfnamefont{R.}~\bibnamefont{Aaij}} \bibnamefont{et~al.}
  (\bibinfo{collaboration}{LHCb}), \bibinfo{journal}{Eur. Phys. J. C}
  \textbf{\bibinfo{volume}{77}}, \bibinfo{pages}{161}
  (\bibinfo{year}{2017}{\natexlab{b}}), \eprint{1612.06764}.

\bibitem[{\citenamefont{Descotes-Genon
  et~al.}(2021)\citenamefont{Descotes-Genon, Novoa-Brunet, and
  Vos}}]{2008.08000}
\bibinfo{author}{\bibfnamefont{S.}~\bibnamefont{Descotes-Genon}},
  \bibinfo{author}{\bibfnamefont{M.}~\bibnamefont{Novoa-Brunet}},
  \bibnamefont{and} \bibinfo{author}{\bibfnamefont{K.~K.} \bibnamefont{Vos}},
  \bibinfo{journal}{JHEP} \textbf{\bibinfo{volume}{02}}, \bibinfo{pages}{129}
  (\bibinfo{year}{2021}), \eprint{2008.08000}.

\end{thebibliography}

\end{document}